\documentclass[%
reprint,
superscriptaddress,
 amsmath,amssymb,
aps,
 prb,
]{revtex4-2}

\usepackage{graphicx}
\graphicspath{{./figs/}}
\usepackage[rgb]{xcolor}
\definecolor{dgreen}{rgb}{0.0,0.7,0.0}
\definecolor{blue}{rgb}{0.0,0.0,1.0}
\definecolor{dred}{rgb}{0.85,0.0,0.0}
\definecolor{black}{rgb}{0.0,0.0,0.0}
\newcommand{\B}{\textcolor{blue}}

\usepackage[unicode, colorlinks=true, allcolors=blue]{hyperref}
\usepackage{braket}
\usepackage[normalem]{ulem}

\DeclareUnicodeCharacter{2212}{-}

\usepackage{dcolumn}
\usepackage{bm}

\usepackage{lipsum}

\usepackage{tabularx}
\newcolumntype{Y}{>{\centering\arraybackslash}X}

\begin{document}

\preprint{APS/123-QED}

\title{Nuclear spin relaxation mediated by donor-bound and free electrons\\in wide CdTe quantum wells}

\author{B.\,F.~Gribakin}
    \email{boris.gribakin@umontpellier.fr}
    \affiliation{Laboratoire Charles Coulomb, UMR 5221 CNRS-Universit\'e de Montpellier, F-34095 Montpellier, France}
    \affiliation{Spin Optics Laboratory, St. Petersburg State University, 198504 St. Petersburg, Russia}    

\author{V.\,M.~Litvyak}
    \affiliation{Spin Optics Laboratory, St. Petersburg State University, 198504 St. Petersburg, Russia}

\author{M.~Kotur}
    \affiliation{Experimentelle Physik 2, Technische Universit\"at Dortmund, 44227 Dortmund, Germany}

\author{R.~André}
    \affiliation{Universit\'e Grenoble Alpes, CNRS, Institut N{\'{e}}el, 38000 Grenoble, France}

\author{M.~Vladimirova}
    \affiliation{Laboratoire Charles Coulomb, UMR 5221 CNRS-Universit\'e de Montpellier, F-34095 Montpellier, France}

\author{D.\,R.~Yakovlev}
    \affiliation{Experimentelle Physik 2, Technische Universit\"at Dortmund, 44227 Dortmund, Germany}
   
\author{K.\,V.~Kavokin}
    \affiliation{Spin Optics Laboratory, St. Petersburg State University, 198504 St. Petersburg, Russia}

\date{\today}

\begin{abstract}
The nuclear spin systems in CdTe/(Cd,Zn)Te and CdTe/(Cd,Mg)Te quantum wells (QW) are studied  using a multistage technique combining optical pumping and Hanle effect-based detection.
The samples demonstrate drastically different nuclear spin dynamics in zero and weak magnetic fields. 
In CdTe/(Cd,Zn)Te, the nuclear spin relaxation time is found to strongly increase with the magnetic field, growing from 3~s in zero field to tens of seconds in a field of 25~G.
In CdTe/(Cd,Mg)Te the relaxation is an order of magnitude slower, and it is field-independent up to at least 70~G.
The differences are attributed to the nuclear spin relaxation being mediated by different kinds of resident electrons in these QWs.
In CdTe/(Cd,Mg)Te, a residual electron gas trapped in the QW largely determines the relaxation dynamics.
In CdTe/(Cd,Zn)Te, the fast relaxation in zero field is due to interaction with localized donor-bound electrons. Nuclear spin diffusion barriers form around neutral donors when the external magnetic field exceeds the local nuclear field, which is about $B_L\approx 0.4$~G in CdTe. This inhibits nuclear spin diffusion towards the donors, slowing down relaxation.
These findings are supported by theoretical modeling. In particular, we show that the formation of the diffusion barrier is made possible by several features specific to CdTe: (i) the large donor binding energy (about $10$~meV), (ii) the low abundance of magnetic isotopes (only $\approx30$\%  of nuclei have nonzero spin), and (iii) the absence of nuclear quadrupole interactions between nuclei.
The two latter properties are also favorable to nuclear spin cooling via optical pumping followed by adiabatic demagnetization.  Under non-optimized conditions we have reached sub-microkelvin nuclear spin temperatures in both samples, lower than all previous results obtained in GaAs.
\end{abstract}

\maketitle

\section{Introduction}
\label{sec:introduction}

The nuclear spin physics of solids has enjoyed a renewed interest from the scientific community in recent years, with modern research being largely stimulated by the possible spintronics and quantum computing applications.
Frequently, nuclear spins are seen as an obstacle to reaching long electron coherence times, especially in quantum dots~\cite{merkulov2002_hyperfineQDs, khaetskii2002_hyperfineQDs}.
However, they can also be made useful.
For example, the feedback from the nuclei can be used for frequency focusing of electron spin precession in quantum dots~\cite{greilich2007sci_hyperfine_QDs}, and there is a growing body of encouraging research on the topic of hybrid quantum registers employing nuclear spins as long-lived quantum bits~\cite{waldherr2014nature, taminiau2014natnano, chekhovich2020natnano, Denning_collective_2019,gangloff_quantum_2019}.
As the nuclear spins do not interact directly with light, they are controlled via hyperfine interaction with electrons, which is typically achieved through resonant or nonresonant optical pumping, with protocols of increasing complexity~\cite{chekhovich2020natnano, gurudev-dutt2007_qubits_diamond, reiner2024high-fi_initialization_4_qubit, bradley2019_10_qubit_register}.

In GaAs, the model semiconductor, every nucleus has spin $I = 3/2$, and therefore possesses a nonzero quadrupolar magnetic moment.
Once the cubic symmetry is lifted by, e.g., strain or defects, the uncontrollable quadrupolar interactions shorten the nuclear spin lifetime and fundamentally limit nuclear spin temperatures achievable via adiabatic demagnetization.
Furthermore, a strong field dependence of nuclear spin relaxation is introduced~\cite{litvyak2021prb, vladimirova2022simultaneous}.
Although there is ongoing research on comparatively strain-free GaAs/(Al,Ga)As quantum dots (see, e.g., Ref.~\cite{chekhovich2020natnano}), the quadrupolar effects can hardly be eliminated completely and become particularly important in small magnetic fields.

Consequently, materials with $I = 1/2$ are intriguing, as the nuclei do possess spin, but completely lack quadrupolar moments, and thus the aforementioned properties of their nuclear spin systems are robust with respect to stress, defects and electric fields.
CdTe is especially interesting in this regard as the local field describing nuclear spin-spin interactions ($B_{L}\approx 0.4$~G \footnote{as determined from the data in Ref.~\cite{nolle1979zpb} following Refs.~\cite{paget1977, litvyak2023prb_spinspin_localfield_GaAs}})  is smaller than in GaAs ($\approx 1$~G~\cite{litvyak2023prb_spinspin_localfield_GaAs}) due to the low abundance of Cd and Te isotopes with nonzero spin (the latter will be referred to as magnetic nuclei hereafter).
This  is expected to reduce the lowest achievable nuclear spin temperature~\cite{AbragamProctor1958,vladimirova2018prb_spinTemp} and to influence nuclear spin diffusion~\cite{khutsishvili1966_nuclear_spin_diffusion}.
Additionally, CdTe is very similar to GaAs in electronic properties, the primary difference being the higher exciton/donor binding energy and, consequently, the smaller Bohr radius $a_{B} \approx 5$~nm~\cite{kheng_negatively_1993,garcia-arellano2019}.

To use the nuclear spin system in a device, one should find a way of controlling it.
A seldom explored possibility consists in using an external magnetic field to manipulate the diffusion barrier~\cite{khutsishvili1966_nuclear_spin_diffusion}.
A nuclear spin diffusion barrier arises when the magnetic field gradient becomes larger than the local field; this hinders spin flips between neighboring nuclei, slowing down and eventually inhibiting nuclear spin diffusion.
Upon applying a threshold magnetic field the bulk polarized nuclei would become separated from the donor site, preventing rapid diffusion and relaxation on the donor after pumping is stopped~\cite{paget1982_spin_diffusion}.
In semiconductors, there is no consensus on the subject of diffusion barriers: although there is some evidence for their formation in GaAs quantum dots at high magnetic fields~\cite{lu2006_diffBarrierGaAs, gong2011_diffBarrierGaAs_theory}, there is more recent evidence to the contrary~\cite{millington-hotze2023natcomm_noDiffBarrierInGaAsQDs}.
Nevertheless, it is tempting to search for diffusion barrier formation in lightly or even unintentionally $n$-doped CdTe, as the relatively strong localization on donors together with the small local field should make the effect quite observable as far as the electrons remain localized.
Conversely, one would not expect to see a significant magnetic field response if the electrons are not strongly localized, disappearing completely in the limit of bulk Korringa relaxation~\cite{korringa1950}.

\begin{figure}[t]
    \centering
   \includegraphics[]{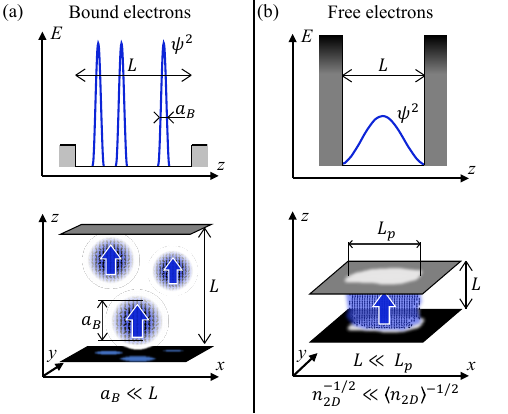}
    \caption{Schematic diagrams of the spin-lattice interactions studied in this work. (a) Donor-bound electrons in a wide insulating CdTe QW with low barriers; nuclear spins interact with the strongly localized individual electron spins. (b) Resident electron gas forming on QW potential fluctuations (shown by barrier color) in a deep CdTe QW with unintentional barrier doping; nuclear spins interact with the weakly localized 2D electron gas. The electron gas puddles \textcolor{black}{are large compared to the QW width, but} constitute only a small portion of the entire QW area ($\langle n_{2D} \rangle$ is the electron gas density averaged over the entire QW).}
    \label{fig:mechanismsCartoon}
\end{figure}

For our study of the CdTe nuclear spin system, we have chosen two illustrative structures.
Both samples are studied in similar conditions via optical pumping followed by adiabatic demagnetization, with photoluminescence (PL) polarization measurements serving as a probe of the nuclear spin state.
Sample~A is a very wide high-quality CdTe quantum well (QW) sandwiched between low (Cd,Zn)Te barriers.
The QW contains a small number of donors due to unintentional {\it n}-doping.
The nuclear spin dynamics develops around these donors\textcolor{black}{, as illustrated in Fig.~\ref{fig:mechanismsCartoon}(a)}.
The inhomogeneous Knight field of the donor-bound electrons creates a diffusion barrier, which results in a magnetic field dependence of the nuclear spin relaxation.
Sample~B is a thinner CdTe QW surrounded by very high (Cd,Mg)Te barriers \textcolor{black}{hosting a dilute two-dimensional (2D) electron gas}. 
\textcolor{black}{However, due to a localization potential along the QW plane formed by well width fluctuations and barrier alloy fluctuations, the electrons are not distributed homogeneously in QW plane. Instead, they are are collected in puddles. The electron gas density inside the puddles ${n_{\rm 2D}}$  is around $5 \times 10^{10}$cm$^{-2}$}, while the rest of the QW remains basically electron-free.
\textcolor{black}{The density of electrons in these puddles is still much greater than the electron density averaged across the entire QW, $n_{\rm 2D}\gg \langle n_{\rm 2D} \rangle$, see Fig.~\ref{fig:mechanismsCartoon}(b)}.
This leads to slow temperature-dependent nuclear spin relaxation induced by fluctuations of the electron spin near the Fermi level in the areas covered by 2D electron gas, with a large contribution due to nuclear spin diffusion from electron-free areas, but with no notable dependence on the applied magnetic field.
Phenomenological modeling of nuclear spin relaxation in these two different regimes supports the proposed interpretations.

The paper is organized as follows.
Section \ref{sec:samples&characterization} describes the studied samples in more detail.
It is followed by Section \ref{sec:experimental} which describes the experimental data obtained.
In Section \ref{sec:discussion} we examine the experimental results and present their theoretical modeling, summarizing our findings in Section \ref{sec:conclusions}.
Finally, Appendices \ref{a.sec:donors} and \ref{a.sec:free} feature extended descriptions of the theoretical models.

\begin{figure}[t]
    \centering
    \includegraphics[]{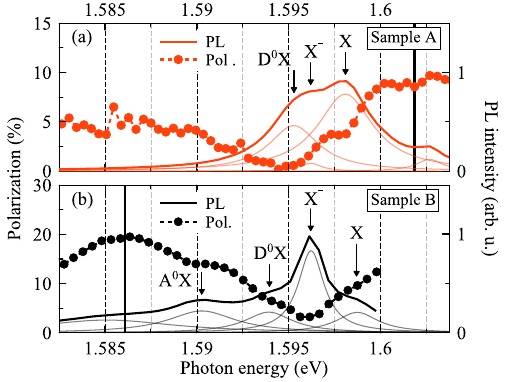}    
    \caption{Low-temperature PL and PL polarization spectra of the two samples studied: (a) Sample~A, (b) Sample~B. Black vertical lines show the energies used for nuclear spin polarization measurements.}
    \label{fig:characterization}
\end{figure}

\section{Samples \& Characterization}
\label{sec:samples&characterization}

The two CdTe QW structures were grown by molecular beam epitaxy.
Sample~A is a $30$-nm-wide CdTe QW sandwiched between Cd$_{0.95}$Zn$_{0.05}$Te barriers, grown on a Cd$_{0.96}$Zn$_{0.04}$Te substrate.
The top barrier also functions as the capping layer and is $93$~nm thick.
\textcolor{black}{Sample~B is a}  $20$-nm CdTe QW surrounded by high Cd$_{0.77}$Mg$_{0.23}$Te barriers. \textcolor{black}{The top barrier (=} cap layer\textcolor{black}{) has a} thickness of  $17.5$~nm.

Both samples are nominally undoped, however, the low-temperature PL spectra shown in Fig.~\ref{fig:characterization} indicate some unintentional doping.
In Sample~A, we interpret the main PL features as the free exciton ground state X ($1.5981$~eV), the negative trion X$^{-}$ ($1.5962$~eV), and the exciton bound to neutral donor D$^{0}$X ($1.5953$~eV), as the binding energies are consistent with literature data~\cite{astakhov2008_CdTe_trion_binding_energy, francou1990prb_CdTe_donors, schmidt1992_CdTe_donors_acceptors_PL}.
The structure of the free exciton resonance series is also reproduced by calculations similar to that of Ref.~\cite{mikhailov2023arXiv}, where a sample grown under the same conditions as Sample~A was studied.

\begin{figure}[t]
    \centering
    \includegraphics{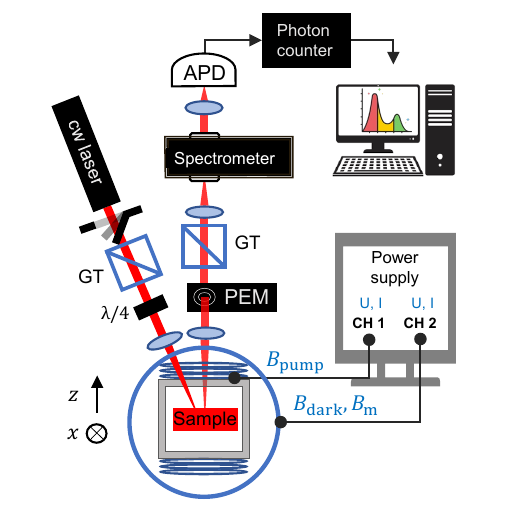}
    \caption{\textcolor{black}{A schematic diagram of the experimental setup. GT are Glan-Taylor prisms, PEM is a photoelastic modulator, APD is an avalanche photodiode. 
    The sample is pumped with circularly polarized light in Faraday geometry {$B_{\rm{pump}}\parallel z $}, and the nuclear spin polarization  is detected via PL polarization degree in Voigt geometry {$B_{\rm{m}}\parallel x $}.}}
    \label{fig:setup}
\end{figure}

In Sample~B the overall blueshift due to additional quantum confinement energy is partly compensated by the enhanced exciton binding energy~\cite{garcia-arellano2019}.
We interpret the spectrum similarly to Sample~A.
Evidently, there are some donors inside the QW as well as acceptors (the D$^{0}$X, A$^{0}$X peaks at $1.5939$ and $1.5903$~eV, respectively)~\cite{schmidt1992_CdTe_donors_acceptors_PL}.
However, the relative intensity of the trion line ($1.5962$~eV) as compared with the free exciton ($1.5988$~eV) suggests that a dilute carrier gas is present.
Although a  structure  similar to Sample~B was shown to contain a hole gas, with the carrier type changing under above-barrier excitation~\cite{bartsch2011_polarization-resolved_trions}, supplementary pump-probe measurements (data not shown) confirm the presence of a \textcolor{black}{resident} electron gas in Sample~B. 
\textcolor{black}{To explain the experimental results in Sec.~\ref{sec:experimental}, we hypothesize that the resident electron gas is inhomogeneous, i.e., there are areas hosting electron gas puddles, and there are areas that are electron-free.}
\textcolor{black}{This electron landscape results from a fluctuating QW potential, which traps electrons coming from the barriers.}
 \textcolor{black}{The roughness of the QW potential could be a consequence of QW width fluctuations and, more probably, barrier height fluctuations, as} due to the height of the Mg barriers, small variations of Mg content lead to rather large QW potential fluctuations.
That is to say, a Mg density variation of just $1\%$ would lead to more than a $10$~meV change in total barrier height~\cite{leblanc2017_CdMgTe_x, adachi2005_semiconductor_properties}.
\textcolor{black}{Overall, the scale of QW potential fluctuations in Sample~B seems to be around several meV, as evidenced by the PL spectrum, and this should be enough to confine the dilute electron gas into puddles at the low temperatures studied, assuming it follows the Fermi-Dirac distribution.}
As we will show in Sec.~\ref{sec:discussion}, the electron gas density in these puddles does not exceed $10^{10}$~cm$^{-2}$.

\textcolor{black}{Both samples are homogeneous on the scale of laser spot size (around $50$~$\mu$m) and show very similar behavior for different pieces taken from the same substrate. 
We have checked this by supplementary measurements of carrier dynamics using photoluminescence and pump-probe Kerr rotation experiments.}

\begin{figure}[t]
    \centering
    \includegraphics{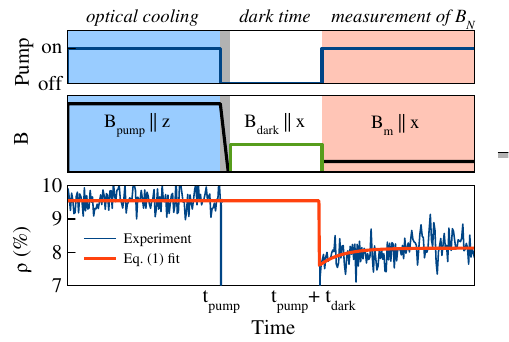}
    \caption{An illustration of the experimental protocol used to measure nuclear spin relaxation as a function of the magnetic field $B_{\rm dark}$. The demagnetization stage (shown by gray area) is very short and takes place between the cooling and relaxation stages. The times $t_{\rm pump}$, $t_{\rm dark}$ are not to scale. The data shown in the lower panel have been measured on Sample~A at $B_{\rm dark} = 25$~G at $t_{\rm dark} = 25$~s.}
    \label{fig:experimentalProtocol}
\end{figure}

\section{Experimental}
\label{sec:experimental}

Nuclear spin polarization experiments were performed following the standard optical orientation technique~\cite{opticalOrientation1984}.
Overall, the setups and experimental protocols are quite similar to Refs.~\cite{kotur2016prb, litvyak2021prb, kotur2021natcomphys, vladimirova2022simultaneous}, and have been extensively used in studies of nuclear spin relaxation in GaAs. 

\textcolor{black}{A schematic diagram of the experimental setup is shown in Fig.~\ref{fig:setup}, while Fig.~\ref{fig:experimentalProtocol} illustrates the experimental protocol.} 
During the optical cooling stage, the nuclear spins are dynamically polarized in a longitudinal magnetic field $B_{\rm pump}$ under optical pumping with circularly polarized light \textcolor{black}{(Faraday geometry)}.
Once the pumping is stopped, we proceed with adiabatic demagnetization to zero field.
During this process, the nuclear spin system reaches internal equilibrium and its lowest spin temperature $\theta_{N}^{(0)}$.
In the next stage, the transverse field $B_{\rm dark}\B{\parallel x}$ is applied, and nuclear spin relaxation proceeds for a time $t_{\rm dark}$, after which we conduct the measurement.

The measurement procedure yields the effective nuclear field $B_{N}$ (the Overhauser field~\cite{overhauser1953_Overhauser_field} that the nuclei exert on the electrons via hyperfine interaction).
The procedure consists in switching on simultaneously the transverse measuring field $B_{\rm m}$ and the pumping beam \textcolor{black}{(Voigt geometry)}, and then recording the PL depolarization.
The PL polarization degree $\rho(t)$ is then determined by the total field $B_{\rm m} + B_{N}$ experienced by the electrons (the Hanle effect), and can be described with
\begin{align}
        \rho(t) = \frac{\rho_{0} B_{1/2}^{2}}{B_{1/2}^{2} \!+\! \left[ B_{\rm m} \!+\! b \!+\! \left(B_{N} (t_{\rm dark}) - b\right) \exp\left( - \frac{t-t_{\rm dark}}{T_{1}}\right) \right]^{2}},
\end{align}
where $\rho_{0}$ is the polarization degree in zero transverse field, $B_{1/2}$ is the half-width of the Hanle curve, $T_{1}$ is the PL polarization relaxation time, and $b$ is a small nuclear field brought about by additional nuclear spin cooling in the Knight field of photocreated electrons during measurement~\cite{litvyak2021prb}.

In experiments on Sample~A, the nuclear spins were pumped for $t_{\rm pump} = 610$~s in an external field of $B_{\rm pump} = 150$~G by a laser tuned to $E = 1.822$~eV (above-barrier excitation), with $B_{\rm m} = 1.2$~G.

Sample~B was subjected to a shorter pumping time of $t_{\rm pump} = 100$~s, and the applied field was also somewhat weaker, $B_{\rm pump} = 115$~G.
The pumping energy was $E = 1.631$~eV (below-barrier excitation), and the measuring field was chosen equal to the pumping field, $B_{\rm m} = 115$~G.
We note that in experiments on both samples the pumping energy was dictated by laser availability, not pumping efficiency.

The rate of spin exchange between, e.g., neutral donors and free electrons is orders of magnitude faster than the nuclear spin dynamics~\cite{paget1982_spin_diffusion}.
Therefore, all of the electron states are equally valid sensors of the nuclear spin state.
As such, the detection energies were chosen to maximize the PL polarization degree $\rho(t)$.
The PL intensity and polarization spectra are shown in Fig.~\ref{fig:characterization}.
Solid black lines indicate the detection energies, 1.6019~eV and 1.5861~eV for Samples A and B respectively.
Additional modeling shows that in Sample~A this roughly corresponds to the free exciton state formed by the e1-hh3 quantum-confined states.
In Sample~B we are likely probing states that are localized on QW potential fluctuations.

\begin{figure}[t]
    \centering
    \includegraphics{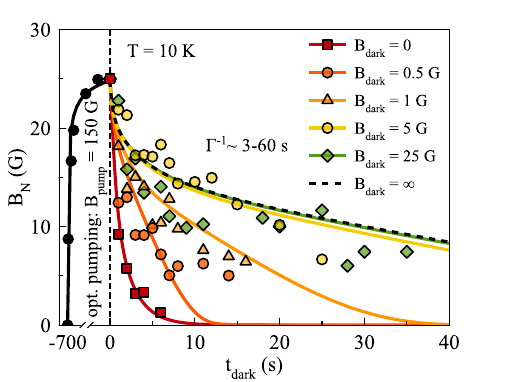}
    \caption{Nuclear spin dynamics in Sample~A. Solid lines: self-consistent theoretical modeling via diffusion equation. The strong dependence on magnetic field is well-described by the onset of a diffusion barrier due to the inhomogeneous Knight field. The curve at $B_{\rm dark} = 0$ corresponds to spatially homogeneous diffusion with $D = 3 \times 10^{-13}$~cm$^{2}$s$^{-1}$ with relaxation via bound electrons. The diffusion barrier becomes more significant as $B_{\rm dark}$ grows, slowing nuclear spin diffusion and impeding relaxation. For details, see Sec.~\ref{subsec:donor-bound} and Appendix~\ref{a.sec:donors}.}
    \label{fig:localizedElectrons}
\end{figure}

\begin{figure}[t]
    \centering
    \includegraphics{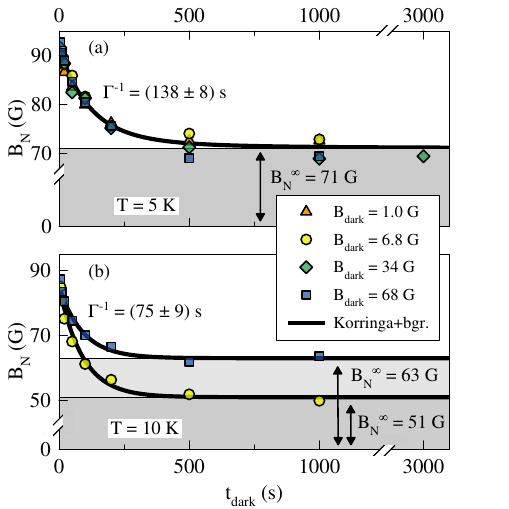}
    \caption{Nuclear spin dynamics in Sample~B at $T=5$~K (a) and 10~K (b). Points: experimental data, solid lines: modeling via Eq.~(\ref{eq:free_e_rate_eq}). The dynamics is governed by free electron-induced relaxation and  diffusion. It is independent of the magnetic field, as the electron localization is insufficient for the diffusion barrier effect. \textcolor{black}{At 5~K essentially nothing changes with the magnetic field.} At 10~K the relaxation rate is  \textcolor{black}{independent of $B_{\rm dark}$}, only the backgrounds are different. For details, see Sec.~\ref{subsec:freeElectrons} and Appendix~\ref{a.sec:free}.}
    \label{fig:freeElectrons}
\end{figure}

The results of measurements on Samples~A and~B are presented in Figs.~\ref{fig:localizedElectrons} and~\ref{fig:freeElectrons}, respectively.
The two samples exhibit drastically different nuclear spin dynamics.

In Sample~A we observe an increase of the nuclear spin lifetime with the transverse magnetic field $B_{\rm dark}$, and it reaches tens of seconds at best.

Sample~B demonstrates a much longer nuclear spin lifetime of the order of hunblacks of seconds, which is independent of $B_{\rm dark}$ at 5~K. 
Additionally, there is a background -- an extremely long-lived component of the nuclear spin polarization that persists for thousands of seconds. 
\textcolor{black}{The nuclear spin relaxation shows a pronounced temperature dependence. 
Measurements at $B_{\rm dark}=6.8$~G and $T=10$~K, see Fig.~\ref{fig:freeElectrons}~(b), indicate the reduction of the rate at $10$~K by almost a factor of $2$ with respect to the rate measured at $5$~K. 
Increasing  $B_{\rm dark}$ up to $68$~G does not affect the relaxation rate. 
By contrast, the long-lived component of the nuclear spin polarization appears to by  sensitive to both temperature and magnetic field.}

It is instructive to estimate the lowest nuclear spin temperature reached in these experiments on CdTe and compare to the existing measurements in GaAs samples.
Since $B_{\rm m} > B_{L}$, from the Curie law we get~\cite{opticalOrientation1984}
\begin{align} 
\label{eq:B_N_Curie}
   \theta_{N}^{(0)} =  \left. \frac{\hbar \langle \gamma  \rangle I(I+1)}{3 k_{B}} \frac{B_{L}b_{N}}{B_N} \right|_{t_{\rm dark}=0},
\end{align}
where $k_{B}$ is  the Boltzmann constant, $\langle \gamma \rangle=-6.47 \times 10^{7}$~rad\,s$^{-1}$\,T$^{-1}$ is the average gyromagnetic ratio of magnetic isotopes~\cite{harris_nmr_2002}, and
\begin{align} 
\label{eq:bn}
  b_{N} = \frac{I\langle A \rangle}{\mu_{B} g_{e}} \approx 700\;{\rm G}
\end{align}
is the value of the nuclear field at saturation, where $\langle A \rangle = -13$~$\mu$eV is the average hyperfine constant~\cite{syperek_long-lived_2011,Testelin_Hole_2009, zhukov2014_NMR_RSA},  $g_{e}=-1.6$ is the electron $g$-factor~\cite{Sprinzl_Influence_2010}, and $\mu_{B}$ is the Bohr magneton.

Applying Eq.~(\ref{eq:B_N_Curie}), we find that the nuclear spins of Sample~A are cooled down to around $0.3$~$\mu$K.
In Sample~B the temperature is even lower, possibly reaching as low as $0.08$~$\mu$K, although it is difficult to make a precise measurement due to the aforementioned in-plane inhomogeneity of Sample~B.
Nevertheless, even after subtracting the background completely, we arrive at around the same nuclear spin temperature as for Sample~A, $\theta_{N}^{(0)}\approx 0.3$~$\mu$K.
For comparison\textcolor{black}{, in GaAs, which is a common platform for dynamic nuclear polarization experiments,} the lowest spin temperatures obtained are $0.5$~$\mu$K (\cite{kotur2021natcomphys}, rotating frame) and $2$~$\mu$K (\cite{vladimirova2018prb_spinTemp}, laboratory frame). 

\textcolor{black}{It is not trivial to compare our results to other materials with $I = 1/2$, due to the lack of available low-field data, and significant differences between implemented experimental approaches.
In principle,  nuclei in silicon are quite interesting because the only magnetic isotope, $^{29}$Si, has only $4.6\%$ natural abundance, and paramagnetic centers allow for efficient dynamic nuclear polarization. In particular, in Ref.~\cite{falk2015prl_SiC_nuclear_spin_temp} a $^{29}$Si spin temperature as low as $5$~$\mu$K is achieved at $300$~G (resonant magnetic field that ensures anti-crossing of electron and nuclear spin levels) near paramagnetic color centers in SiC.  In Ref.~\cite{mccamey2009prl_P_in_Si_DNP},  spin temperature of $5$~mK for phosphorus nuclei spin subsystem in Si:P is reported at $8.6$~T.  This result relies on electrically detected magnetic resonance, and requires high fields and an electric current. In both cases, however, direct comparison with the zero-field temperature determined is this work is difficult.}

Considering that \textcolor{black}{in our experiments} such a low temperature is obtained under pumping conditions which were not optimized for the most efficient nuclear spin cooling, these results suggest that CdTe is a potential candidate for reaching record low temperatures in the semiconductor nuclear spin system.


\section{Discussion}
\label{sec:discussion}

\subsection{Donor-bound electrons (Sample~A)}
\label{subsec:donor-bound}

\begin{figure}[t]
    \centering
    \includegraphics{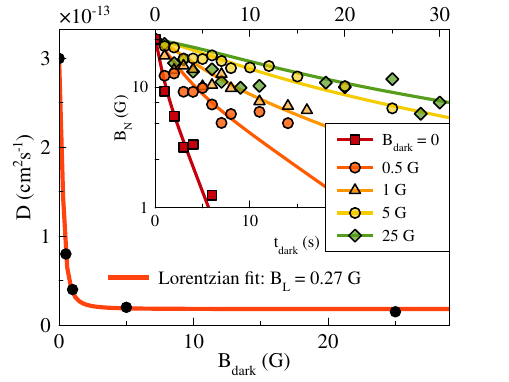}
    \caption{
    Results of modeling the Sample A nuclear spin dynamics with a field-dependent diffusion coefficient $D(B_{\rm dark})$. Sharp changes in diffusion occur in fields close to the local field $B_{L} \approx 0.2-0.4$~G. Inset: corresponding fits of the data.}
    \label{fig:localizedElectrons_D(B)}
\end{figure}

The nuclear spin relaxation in Sample~A (see Fig.~\ref{fig:localizedElectrons}) is rather fast, with nuclear spin lifetimes of the order of seconds.
We assume that the relaxation is due to hyperfine interaction with donor-bound electrons, as these are essentially the only resident electrons in the sample.
The distinct field-dependent behavior that we observe can be phenomenologically modeled by a radial diffusion equation with a nuclear spin diffusion coefficient $D$ that depends on $B_{\rm dark}$, see Fig.~\ref{fig:localizedElectrons_D(B)}.
To account for the experimental results, the diffusion coefficient must experience a sharp change in fields of the order of $0.3$~G, close to the value of the local field $B_{L}$.
We theoretically estimate the local field due to both magnetic dipole-dipole interaction and indirect interactions to be $\lesssim 0.4$~G.
This is consistent with NMR linewidth measurements in CdTe, which give a value of about $0.2$~G for $^{125}$Te~\cite{nolle1979zpb}.
The field dependence of the diffusion coefficient suggests the formation of a nuclear spin diffusion barrier.
Below we briefly explain the principles of nuclear spin relaxation affected by the diffusion barrier; further details can be found in Appendix~\ref{a.sec:donors}.

In weak magnetic fields $B_{\rm dark} \gtrsim B_{L}$, the cold nuclear spins are easily polarized, resulting in an Overhauser field \textcolor{black}{$B_{N} \gg B_{\rm dark}$}.
The polarized nuclei interact with electrons.
Since the resident electrons are localized on donors and their density is low, the hyperfine relaxation mechanism is dominant over the spin-orbit one.
It can be shown that in this regime dynamic electron polarization by nuclei takes place~\cite{vladimirova2021_polaron}, \textcolor{black}{which means that} the average electron spin $\langle S_{z} \rangle$ is determined by the mean nuclear spin $\langle I_{z} \rangle$, and is therefore proportional to the Overhauser field $B_{N}$.
In turn, the polarized electrons exert a weak but highly inhomogeneous Knight field back upon the nuclei, $B_{e}({\bf r}_{j}) \propto |\psi({\bf r}_{j})|^2 \langle S \rangle$, where $\psi({\bf r}_{j})$ is the electron wavefunction at the nucleus located at ${\bf r}_{j}$.
Indeed, the Knight fields experienced by two nearby nuclei situated at a distance $r$ away from the donor and separated by $r_{ij} \ll a_{B}$ differ by: 
\begin{align} \label{eq:DeltaB_e(r)}
    \Delta B_{e}(r , B_{N}) = r_{ij} \left| \frac{\partial B_e(r)}{\partial r} \right|.
\end{align} 
If at any distance $r$ from the donor $\Delta B_{e}(r, B_{N}) > B_{L}$, then the corresponding nuclear spin flips are suppressed due to energy conservation.
Eq.~(\ref{eq:DeltaB_e(r)}) allows us to estimate that at $B_{\rm dark}$ as small as $1$~G  in the immediate vicinity of the donor site $\Delta B_{e}(r, B_{N}) > B_{L}$ even for a pair of nearest-neighbor nuclei with nonzero spin separated by $r_{\rm nn}$, so nuclear \textcolor{black}{spin} diffusion is expected to be strongly suppressed.

The rigorous calculation of the spatial and field-dependent behavior of the diffusion coefficient is outside the scope of this work.
Instead, we incorporate the effect by \textcolor{black}{introducing a phenomenological dependence of} the diffusion coefficient $D$ \textcolor{black}{on} the Overhauser field $B_{N}$ through \textcolor{black}{the Knight field gradient} $\Delta B_{e}(r_{\rm nn}, B_{N})$ (hereafter referred to as $\Delta B_{e}^{\rm nn}$) as defined by Eq.~(\ref{eq:DeltaB_e(r)}).
This \textcolor{black}{dependence must ensure that}
\begin{align}
    \qquad D(\Delta B_{e}^{\rm nn}) &\approx 0, && {\rm if} &&& \Delta B_{e}^{\rm nn} \gg B_{L}; \label{eq:D_asymptotics1} \qquad\\
    \qquad D(\Delta B_{e}^{\rm nn}) &\approx D_{0}, &&{\rm if}  &&& \Delta B_{e}^{\rm nn} \ll B_{L}. \label{eq:D_asymptotics2} \qquad
\end{align}
In the above, $D_{0}$ is the nuclear spin diffusion coefficient in the absence of Knight-field feedback, i.e., in zero external field.
Thus, \textcolor{black}{the nuclear spin diffusion barrier builds up wherever the Overhauser field induces a sufficiently inhomogeneous Knight field}.
In the theoretical curves of Fig.~\ref{fig:localizedElectrons}, a step-like $D(\Delta B_{e}^{\rm nn})$ was used, accurately describing the data with $B_{L} = 0.4$~G as calculated from the spin-spin interaction constants measured by Nolle~\cite{nolle1979zpb, litvyak2023prb_spinspin_localfield_GaAs}.
Additional details regarding the modeling can be found in Appendix~\ref{a.sec:donors}.

Note that the optical cooling stage, where the strong field $B_{\rm pump}$ is applied, is modeled with the diffusion coefficient $D = 0$, while the curve at $B_{\rm dark} = 0$ corresponds to the maximum diffusion coefficient $D_{0} = 3\times 10^{-13}$ cm$^{2}$s$^{-1}$, which is around what is typically seen in GaAs~\cite{paget1982_spin_diffusion, vladimirova2017_nuclear_spin_relaxation_nGaAs}.

\subsection{Free electrons (Sample~B)}
\label{subsec:freeElectrons}

As discussed in the previous Section, the seconds-long nuclear spin relaxation and its magnetic field dependence in Sample~A are direct consequences of strong electron localization.
Since in Sample~B we observe nuclear spin relaxation times of the order of a hunblack seconds, the electrons responsible for this relaxation must be much less localized.
Indeed, as discussed in Sec.~\ref{sec:samples&characterization},  \textcolor{black}{there is a resident electron gas in Sample~B}, so we turn our attention to free 2D electrons.

In structures hosting an electron gas, nuclear spin relaxation via interaction with free electrons is known to be relevant~\cite{kotur2014korringa, vladimirova2017_nuclear_spin_relaxation_nGaAs, tracy2006korringa_landauLevels, evers2019_CdTeQW_modulationDoped}.
In the extreme case of a degenerate electron gas, one expects the Korringa relaxation mechanism to dominate.
Its defining feature is the simple relationship $\gamma_{\rm K} \propto k_B T$ between the Korringa relaxation rate $\gamma_{\rm K}$ and the temperature of the electron gas~\cite{korringa1950}.
This is consistent with our measurements, which yield $\Gamma^{-1} = (138 \pm 8)$\,s at $T = 5$~K and $\Gamma^{-1} = (75 \pm 9)$\,s at $T = 10$~K, as determined from single-exponential fits of the data shown in Fig.~\ref{fig:freeElectrons}.
\textcolor{black}{Moreover, this excludes localized electrons as the source of nuclear spin relaxation, as one would expect the opposite trend with increasing temperature~\footnote{Delocalization of electrons with increasing temperature would lead to reduced hyperfine interaction, which would slow down nuclear spin relaxation, in contrast to our observations.}.}
Thus, we continue with free electron-mediated relaxation.

\textcolor{black}{To calculate the nuclear spin relaxation rate due to interaction with a dilute electron gas in a QW}, we employ an approach similar to Ref.~\cite{berg1990korringa}. 
\textcolor{black}{That is, we calculate the spin-flip matrix elements due to contact hyperfine interaction, average them over the nuclear spins under the electron orbit, then calculate their average considering the Fermi distribution of occupied and free electron states; the application of Fermi's golden rule completes the calculation.}
\textcolor{black}{The complete calculation is reported in Appendix~\ref{a.sec:free}. Here we only present the main result, which is the nuclear spin relaxation rate that accounts for a low-density 2D electron gas:
\begin{align} \label{eq:korringa_rel_rate}
    \Gamma_{\rm K}(z) =& \,\frac{\pi}{\hbar} \frac{k_{B} T}{1 + e^{-E_{\rm F} / k_{B} T}} \langle A^{2} \rangle \, \nu_{0}^{2} \, g_{\rm 2D}^{2} |\psi_{e, \rm free} (z)|^{4}.
\end{align}
In the above, $\langle A^{2}\rangle$ is the square of the hyperfine constant averaged over the magnetic isotopes in CdTe, $\nu_{0}$ is the elementary cell volume, $E_{\rm F}$ is the Fermi energy, $g_{\rm 2D}$ is the 2D electron density of states, and $\psi_{e, {\rm free}}(z)$ is the cosine-like electron wavefunction. 
The result is similar to the usual bulk Korringa relaxation~\cite{korringa1950} and the results of Ref.~\cite{berg1990korringa}.} 
However, we do obtain a low-density correction which becomes significant if the Fermi energy is small compared to $k_{B} T$, i.e., when the gas is nondegenerate.
As the Korringa mechanism is reserved for degenerate electron gases, hereafter we will refer to the resulting relaxation as Korringa-{\it like}.

A prominent feature of nuclear spin relaxation via 2D electrons is the dependence of their relaxation rate $\Gamma_{\rm K}$ on the position of the nucleus along the QW growth axis:  $\Gamma_{\rm K} \propto \ |\psi_{e, {\rm free}}(z)|^{4}$.
 Performing a calculation of $\Delta B_{e}^{\rm nn}$ similar to that made for Sample~A via Eq.~(\ref{eq:DeltaB_e(r)}), we find that during pumping, when the Knight field is at its maximum due to the high density and average spin \textcolor{black}{of photocreated electrons}, a soft diffusion barrier may arise along the growth axis.
During relaxation, however, confinement-induced localization \textcolor{black}{of resident electrons} in the QW is certainly insufficient to significantly affect spin diffusion along $z$, and the diffusion barrier cannot be formed.

The large long-lived backgrounds seen in Fig.~\ref{fig:freeElectrons} are quite surprising, especially in comparison with Sample~A.
We believe they can only be explained by \textcolor{black}{assuming that} the entirety of the QW \textcolor{black}{is} polarized during pumping\textcolor{black}{, while the resident electron gas is only present in some areas of the QW, forming puddles}.
\textcolor{black}{The former} is consistent with Ref.~\cite{leung2004NMR_CdTe_bulk_polarization}, where homogeneous polarization occurs under resonant pumping.
Although in our case the optical pumping is nonresonant, Sample~B is pumped below the barrier, which ensures efficient polarization of the QW nuclei.
After pumping we observe a slow relaxation process (see Fig.~\ref{fig:freeElectrons}), which could be caused by two factors.
First of all, the unpolarized resident electron gas siphons the nuclear spin polarization via Korringa-like relaxation as discussed above.
On the other hand, it is possible that simultaneously a portion of the polarization escapes into the electron-free parts the QW via spin diffusion.
However, the resident electrons induce relaxation also during pumping.
Therefore, the nuclei situated within electron puddles should not be more polarized than the rest of the QW, and the second mechanism is unlikely to be significant.
Eventually, a quasi-stationary regime is reached, where the flow of nuclear spin into the puddles becomes completely compensated by the slow Korringa-like relaxation, leading to the long-lived signal.

It is not feasible to meaningfully model the experiments on Sample~B with a diffusion equation similarly to Sec.~\ref{subsec:donor-bound}, as that requires the knowledge of the in-plane electron \textcolor{black}{gas} localization area \textcolor{black}{(i.e., the puddle size)} in addition to other parameters such as the Fermi energy and the diffusion coefficient.
To evade this, we spatially integrate the diffusion equation over the area occupied by the resident electron gas.
This yields a rate equation for the Overhauser field, or, equivalently, for the average nuclear spin $I_{z}= I B_{N} / b_{N}$ with the average Korringa-like relaxation rate $\Gamma_{\rm K}^{\rm (avg)} = L^{-1} \int_{\rm QW} \Gamma_{\rm K}(z) {\rm d}z$.
An additional term $P_{D}(t)$ describes the \textcolor{black}{nuclear spin} diffusion flow into or out of the area:
\begin{align} \label{eq:free_e_rate_eq}
    \frac{\partial I_{z}(t)}{\partial t} = P_{D}(t) - \Gamma_{\rm K}^{\rm (avg)} I_{z}(t).
\end{align}
Because one cannot easily calculate $P_{D}(t)$, we use a simplified phenomenological expression:
\begin{align} \label{eq:free_e_rate_eq_P_D_t}
    P_{D}(t) = \Gamma_{D} \left[J_{z} - I_{z}  (z)\right],
\end{align}
where $J_{z}$ is the average nuclear spin of nuclei unaffected by Korringa-like relaxation, and $\Gamma_{D}$ is the diffusion rate.
The diffusion coefficient cannot be extracted with this approach, unfortunately.
Furthermore, only the sum $\Gamma_{\rm K}^{\rm (avg)} + \Gamma_{D}$ is determined from experiments, although we would expect $\Gamma_{D} \ll \Gamma_{\rm K}^{\rm (avg)}$, as discussed above.

Fitting the model defined by Eqs.~(\ref{eq:free_e_rate_eq}-\ref{eq:free_e_rate_eq_P_D_t}) to the data enables us to give the lower limit of the nuclear spin polarization of the entire QW in Sample~B, which is around $15 \%$ at $T=5$~K and $6.5-8 \%$ at 10~K.
These estimates are obtained under the assumption that the observed relaxation is entirely due to diffusion, so $\Gamma_{D} \gg \Gamma_{\rm K}^{\rm (avg)}$, which is unlikely.
Therefore, we expect that the polarization of the electron-free areas is much higher in reality.
On the other hand, if all of the relaxation was due to the resident electron gas, that would require \textcolor{black}{small} 2D densit\textcolor{black}{ies} of approximately $5\times10^{9}$~cm$^{-2}$ and $8\times10^{9}$~cm$^{-2}$ for $T = 5$~K and $10$~K, respectively (see Fig.~\ref{fig:korringa_vs_density} in the Appendix), which are reasonable carrier densities considering the PL spectrum in Fig.~\ref{fig:characterization}(b)~\cite{bartsch2011_polarization-resolved_trions, zhukov2014_NMR_RSA}.
The difference between the two densities is also unsurprising, as it is possible that at higher temperatures more electrons from the barriers are collected into the QW, and more donor-acceptor pairs inside the QW are ionized.
The noticeable difference between the backgrounds in $B_{\rm dark} = 6.8$~G and $68$~G data at $T = 10$~K is not clear as the nature of the background signal is not exactly identified.
Additional experiments are required to elucidate this issue.

\section{Conclusions}
\label{sec:conclusions}

The nuclear spin dynamics in CdTe QWs is shown to depend strongly on the nature of the electrons present in the sample.

In the presence of donor-bound electrons, like in the CdTe/(Cd,Zn)Te QW studied here, the enhanced role of the  hyperfine interaction is a direct consequence of features specific to CdTe: the low density of magnetic nuclei and absence of quadrupole nuclear moments lead to very efficient nuclear spin cooling, while the large binding energy of electrons produces a strong hyperfine interaction.
Thus, the normally weak effect of dynamic polarization of electron spins by the nuclei is enhanced, and leads to the formation of a nuclear spin diffusion barrier.
The nuclear spin relaxation is then easily tunable.
 In zero field nuclear spin diffusion is efficient and the relaxation time is of the order of seconds.
In fields larger than $B_{L}$, diffusion is strongly suppressed, increasing the nuclear spin lifetime by more than an order of magnitude in a field of 25~G. A model that accounts for the spatial inhomogeneity of the Knight field and the resulting field-dependent diffusion barrier allows us to quantitatively describe the experimental results.

On the other hand, if resident electrons form a nondegenerate 2D electron gas, as it is in the CdTe/(Cd,Mg)Te QW studied in this work, nuclear spin relaxation is much slower, and is field-independent at least up to fields of the order of $70$~G.
At $T = 10$~K, the relaxation time is of the order of $70$~s, and two times slower at $T = 5$~K, much longer than \textcolor{black}{what was} measured \textcolor{black}{near neutral donors} in the CdTe/(Cd,Zn)Te QW.

Notably, not only the nuclei in the areas covered by electron gas puddles, but the nuclear spin system across the entire QW plane is polarized during below-barrier pumping.
This introduces a spin diffusion flow into the areas occupied by the resident electrons.
Eventually a steady state is reached, where we \textcolor{black}{observe} a stable nuclear spin polarization that persists for thousands of seconds, dynamically supported by spin diffusion.
Modeling shows that at least a $\sim15\%$ polarization of the bulk nuclei is achieved at $T = 5$~K, although further study is required to completely unveil the nature of the long-lived signal and its temperature dependence.

Our theoretical description of free electron-mediated relaxation includes a low-density correction, which generalizes the Korringa relaxation mechanism for arbitrary QW electron densities.
This may become useful if the electron gas is purposely injected, either electrically or by modulation doping, at densities below the Mott transition.

Finally, the sub-microkelvin temperatures that we reach by unoptimized optical pumping indicate that CdTe is a potential candidate for reaching record low nuclear spin temperatures, presumably due to the low abundance of magnetic isotopes and their zero quadrupole moment.

\section{Acknowledgements}
\label{sec:acknowledgements}

The authors are grateful to S.~Cronenberger and D.~Scalbert for helpful and valuable discussions.
BG and MV acknowledge financial support from French National Research Agency, Grant No. ANR-21-CE30-0049 (CONUS).
BG acknowledges support from the French Embassy in Moscow (Vernadskii fellowship for young researchers 2021).
Contributions of MK and DY are supported by the Deutsche Forschungsgemeinschaft within the International Collaborative Research Center TRR 160 (project A6).
VL and KK acknowledge financial support from Russian Science Foundation, Grant No. 22-42-09020.
RA has benefited from the technical and scientific environment of the CEA-CNRS joint team "Nanophysics and Semiconductors".

\appendix 

\section{Nuclear spin relaxation and diffusion in the vicinity of donor-bound electrons}
\label{a.sec:donors}

We describe the nuclear spin system in CdTe by the average nuclear spin $\langle I_{z}({\bf r}, t) \rangle$ as experienced by the electrons, so the angled brackets imply a spatial average over an area that is small in comparison to $a_{B}$, but large enough to include thousands of nuclei.
To model the behavior of the nuclear spin polarization, following the seminal work of D.~Paget~\cite{paget1982_spin_diffusion} and many others, we write a spin diffusion equation with relaxation and pumping terms.
In general, this equation has the form
\begin{align} \label{a.eq:diffusion_diffEq}
    \frac{\partial \langle I_{z} \rangle}{\partial t} =& \nabla \left[ D({\bf r}, t | B) \nabla \langle I_{z} \rangle \right] - \Gamma({\bf r}, t) \langle I_{z} \rangle + P({\bf r}, t).
\end{align}
Here $\Gamma({\bf r}, t)$ and $P({\bf r}, t)$ are the electron-mediated nuclear spin relaxation and polarization rates, and $D({\bf r}, t | B)$ is the magnetic field-dependent diffusion coefficient, which generally is inhomogeneous and time-dependent.
The experimentally measured Overhauser field acting on the donor-bound electron is recovered by performing the integration
\begin{align}
    B_{N}(t) = \frac{b_{N}}{I} \int \langle I_{z}({\bf r}, t) \rangle |\psi_{e, {\rm loc}}({\bf r})|^2 {\rm d} {\bf r},
\end{align}
where $\psi_{e, {\rm loc}}(r) = (\pi a_{B}^{3})^{-1/2} \exp \left( -r / a_{B} \right)$ is the wavefunction of the donor-bound electrons.

Let us consider a perfect crystal that is weakly doped with donors.
That is to say, at low temperatures the only electrons in the system are the electrons localized on donors, which are separated by distances much greater than the Bohr radius ($a_{B}\approx 5$~nm in CdTe).
Since all electrons contribute to the signal independently, it is sufficient to describe what happens under one donor orbit.
The spherical symmetry of the problem reduces Eq.~(\ref{a.eq:diffusion_diffEq}) to its radial part.

Because of the above-barrier pumping and the strong localization of a donor-bound electron, we assume that both the pumping and the relaxation occur via the donor-bound electrons.
It is possible to directly calculate the nuclear spin pumping and relaxation rates due to spin-flips with a donor-bound electron described by the wavefunction $\psi_{e, {\rm loc}}(r)$.
However, here we limit ourselves to a simpler approach.
The nature of the contact hyperfine interaction requires the rates to be proportional to $|\psi_{e, {\rm loc}}(r)|^{4}$.
We may therefore simply write
\begin{gather}
     \Gamma(r, t) = \Gamma_{0} \exp\left( -4r/a_{B} \right),\\
     P(r, t) = P_{0} \exp\left( -4r/a_{B} \right),
\end{gather}
with $\Gamma_{0}$ and $P_{0}$ being the relaxation and pumping rates at the center of the donor.
In the steady-state, these rates are related through~\cite{paget1977, abragam1961}
\begin{align} \label{a.eq:p_e}
    P(r, t) = \frac{I+1}{S+1} \Gamma(r, t) p_{e},
\end{align}
where $S = 1/2$ is the electron spin,  $I = 1/2$ is the spin of magnetic isotopes in CdTe, and $p_{e} = \langle S_{z} \rangle / S$ is the electron spin polarization.
The fitted curves in Fig.~\ref{fig:localizedElectrons} correspond to $P_{0} = 0.54$~s$^{-1}$ and $\Gamma_{0} = 13.2$~s$^{-1}$ during pumping ($p_{e} = 0.04$), and $\Gamma_{0} = 0.08$~s$^{-1}$ during relaxation.

\begin{figure}[t]
    \centering
    \includegraphics[]{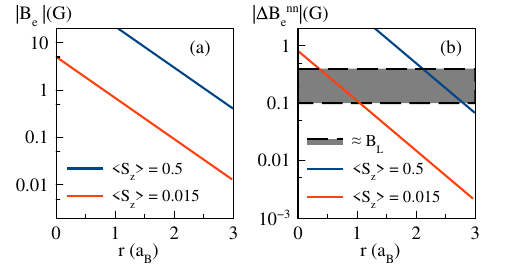}
    \caption{(a) Magnitude of the Knight field produced by a donor-bound electron vs. distance from donor $r$, measured in units of $a_B = 5$~nm; (b) difference of Knight fields experienced by two nuclei separated by the most probable nearest-neighbor distance $r_{\rm nn} = 0.41$~nm. Colors denote values of electron polarization: $\langle S_{z} \rangle = 0.5$ (maximum value~\cite{opticalOrientation1984}) and $\langle S_{z} \rangle = 0.015$ (estimated from experimental nuclear polarization right after pumping, $B = 1$ G). Gray area in (b): approximate value of local field $B_{L}$.}
    \label{fig:diffusionBarrierKnight}
\end{figure}

As stated in the main text, the field dependent behavior of nuclear spin relaxation observed in Sample~A can be understood in terms of a diffusion barrier that builds up due to the inhomogeneity of the Knight field.
As the cold nuclear spins are polarized by a magnetic field $B > B_{L}$, the electrons are polarized via mutual spin-flips with nuclei~\cite{vladimirova2021_polaron}, creating a strongly inhomogeneous Knight field $B_{e}(r, t) \sim \exp(-2r / a_{B})$.
If $B_{e}$ is so strong that the Zeeman splitting of neighboring nuclear spins in $B_{e}$ is greater than in the local field $B_{L}$, then mutual spin-flips are forbidden by energy conservation, and diffusion is hindered.

The distance $r_{\rm nn}$ between two nearest-neighbor magnetic nuclei may be estimated assuming a continuous Poisson distribution with an average density of magnetic isotopes per unit volume $\sigma$:
\begin{align}
    r_{\rm nn} = \int_{0}^{\infty} r e^{-\frac{4}{3} \pi \sigma r^{3}} 4\pi r^{2} {\rm d} r \approx 0.55 \, \sigma^{-1/3}.
\end{align}
Taking into account the isotopic abundances from Table~\ref{table:parameters} along with the CdTe lattice constant  $a_{0} = 0.648$~nm, we obtain $r_{\rm nn} = 0.41$~nm.
The Knight field experienced by an average magnetic nucleus may be written as:
\begin{align}
      B_{e}({\bf r}) =-{\nu_{0}} \frac{\sum_{\alpha} x_{\alpha} A_{\alpha} / \gamma_{\alpha}}{\sum_{\alpha} x_{\alpha}}|\psi_{e, {\rm loc}}(r)|^2 \langle S_{z} \rangle,
\end{align}
where $A_{\alpha}$, $\gamma_{\alpha}$ and $x_{\alpha}$ are the hyperfine constants, gyromagnetic ratios and abundances of magnetic isotopes in CdTe (see Table \ref{table:parameters}), and $\nu_0 = a_{0}^{3} / 4 $ is the elementary cell volume.
The average electron spin is tied to the Overhauser field via
\begin{align} \label{a.eq:avg_I_S_ratio}
\langle I_{z} \rangle = \langle S_{z} \rangle \frac{I(I+1)}{S(S+1)},
\end{align}
which is another form of Eq.~(\ref{a.eq:p_e}).
The difference of Knight fields experienced by two nuclei is given by Eq.~(\ref{eq:DeltaB_e(r)}).
A simple calculation of $\Delta B_{e}^{\rm nn}$ is displayed in Fig.~\ref{fig:diffusionBarrierKnight}, showing that a diffusion barrier should indeed arise at distances of the order of $a_{B}$, depending on the electron spin polarization.

\begin{figure}[t]
    \centering
    \includegraphics{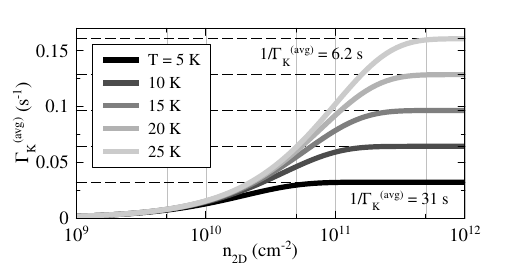}
    \caption{Average Korringa-like relaxation rate $\langle \Gamma_{\rm K} \rangle$ as a function of 2D electron gas density at various temperatures in a 20-nm CdTe QW, calculated according to Eqs.~(\ref{a.eq:korringa_rel_rate}-\ref{a.eq:density_formula}). The relaxation time of a degenerate electron gas $\tau_{\rm K}^{\rm d}$ is marked for $T = 5$~K and $25$~K, highlighting the distinct $\tau_{\rm K}^{\rm d} T = \rm const$ behavior.}
    \label{fig:korringa_vs_density}
\end{figure}

We now suggest a phenomenological expression to model $D({\bf r}, t | B)$.
The asymptotic conditions of Eqs.~(\ref{eq:D_asymptotics1}-\ref{eq:D_asymptotics2}) can be fulfilled most easily by a step function
\begin{gather} \label{a.eq:KnightGradient_diffusionCoefficient}   
    D({\bf r}, t | B) = D_{0} \ H(B_{L}^{\rm (fit)} - \Delta B_{e}^{\rm nn}),
\end{gather}
where $H(x)$ is the Heaviside function, and $B_{L}^{\rm (fit)}$ is a fitting parameter close to the value of $B_{L}$.
Of course, the choice of fitting function is not unique.
We have also tested Lorentzian, Gaussian and Fermi functions, as well as a smoothed Heaviside function with similar results, so Eq.~(\ref{a.eq:KnightGradient_diffusionCoefficient}) was chosen for simplicity.

A problem that is inherent in this approach has to do with the Knight field following the evolution of the nuclear spin polarization.
Because of this, when solving the equation (\ref{a.eq:diffusion_diffEq}), in principle one must recalculate $B_{N} \sim \int \langle I_{z}(r, t) \rangle |\psi_{e, \rm loc}(r)|^{2} r^{2} {\rm d}r$ at every time-step of the calculation.
We evade this by first modeling the experimental data with Eq.~(\ref{a.eq:diffusion_diffEq}) using a set of constant diffusion coefficients $D(B_{\rm dark})$, as shown in Fig.~\ref{fig:localizedElectrons_D(B)}.
The resulting theoretical curves are used as estimates of the Overhauser field for the purposes of calculating the diffusion coefficient according to Eq.~(\ref{a.eq:KnightGradient_diffusionCoefficient}).
Injecting these into Eq.~(\ref{a.eq:diffusion_diffEq}), we obtain the theoretical curves in Fig.~\ref{fig:localizedElectrons}, which display a quantitative agreement with the experimental observations.

\begin{table}[h]
  \caption{Isotope parameters used throughout the work. $x_{\alpha}$ are the natural abundances, $A_{\alpha}$ are the hyperfine constants, and $\gamma_{\alpha}$ are the nuclear gyromagnetic ratios. Throughout this work, we neglect the $^{123}$Te isotope, as its natural abundance is less than 1\%. Data are taken from Refs.~\cite{zhukov2014_NMR_RSA,syperek_long-lived_2011}.}
  \label{table:parameters}
  \renewcommand{\arraystretch}{1.25}
  \begin{tabular}{>{\centering\arraybackslash}p{0.18\columnwidth}>{\centering\arraybackslash}p{0.18\columnwidth}>{\centering\arraybackslash}p{0.18\columnwidth}>{\centering\arraybackslash}p{0.36\columnwidth}}
        \hline\hline
    Isotope & {$x_{\alpha}$} & {$A_{\alpha}$~($\mu$eV)} & {$\gamma_{\alpha}$~($10^{7}$\,rad\,s$^{-1}$\,T$^{-1}$)} \\ \hline
    $^{111}$Cd & 0.13                             & -37.4                            & -5.698                       \\
    $^{113}$Cd & 0.12                             & -39.1                            & -5.960                       \\
    $^{125}$Te     & 0.07                             & -45                              & -7.059                       \\ \hline\hline
  \end{tabular}
\end{table}

\section{Nuclear spin relaxation induced by hyperfine relaxation with 2D electron gas}
\label{a.sec:free}

In this Section, we describe the calculation of the Korringa-like nuclear spin relaxation rate $\Gamma_{\rm K}$ introduced in Sec.~\ref{subsec:freeElectrons}.
Let us consider a perfect infinite QW of width $L$ with an area $S$ hosting a 2D electron gas described by Fermi statistics.
Applying the same approach as in Ref.~\cite{berg1990korringa}, we calculate the matrix elements of the electron-nucleus spin-flip transitions due to hyperfine interaction.

Let $S_{z}$ be the electron spin projection on the QW growth axis, and let $I_{z, j}$ be that for the $j$-th nucleus.
The free electron's wavefunction in the QW $\psi_{e, \rm free}({\bf r})$ is given by:
\begin{align}
\psi_{e, \rm free}({\bf r}) = \sqrt{\frac{2}{LS}} \cos \left( \frac{\pi}{L}z\right).
\end{align}
Denoting the hyperfine interaction operator as $\hat{V}_{j}({\bf r}) = A_{j} \, \nu_{0} \, \hat{\bf I}_{j} \!\cdot\! \hat{\bf S} \, \delta({\bf r} - {\bf r}_{j})$, where  $\hat{\bf S}$ and $\hat{\bf I}_{j}$ are the spin operators for electron and $j$-th nucleus, respectively, we may write the matrix elements for the mutual spin-flips of the electron and a nucleus as
\begin{align}
    M_{\pm, j} = \bra{S_{z} \mp 1, I_{z,j} \pm 1, \psi_{e, \rm free}} \hat{V}_{j} \ket{S_{z}, I_{z,j} \psi_{e, \rm free}}.
\end{align}
A straightforward calculation gives
\begin{align}
    \left| M_{\pm, j} \right|^{2} =& \, \frac{1}{4}\left(A_{j} \, \nu_{0}\right)^{2} \left| \psi_{e, \rm free}(z_{j}) \right|^{4} \nonumber\\\
    &\times \left[ I(I+1) - I_{z, j}^{2} \mp I_{z, j} \right].
\end{align}

To deal with the average nuclear spin $\langle I_{z}({\bf r}, t) \rangle$ interacting with the electron at a given position, we average the matrix elements as 
\begin{align}
    \left| M_{\pm}(z) \right|^{2} =& \frac{1}{4} \langle A^{2} \rangle \, \nu_{0}^{2} \left| \psi_{e, \rm free}(z) \right|^{4} \nonumber\\\
    &\times \left[ I(I+1) - \langle I_{z}^{2} \rangle \mp \langle I_{z} \rangle \right].
\end{align}
The average square of the hyperfine constant $\langle A^{2} \rangle$ is given by
\begin{align} \label{a.eq:<A^2>}
    \langle A^{2} \rangle = \frac{\sum_{\alpha} A^{2}_{\alpha} x_{\alpha}}{\sum_{\alpha} x_{\alpha}},
\end{align}
where the sum goes over all magnetic isotopes in CdTe, see Table~\ref{table:parameters}.

To obtain the nuclear spin flip rate we average the matrix elements over the statistics of the electron gas, i.e., the Fermi-Dirac distribution $f(E, E_{\rm F}) = \left( 1 + \exp \left[(E - E_{\rm F}) / k_{B} T\right] \right)^{-1}$.
According to the Fermi's golden rule
\begin{align} \label{a.eq:flipRate}
    W_{\pm} =& \frac{2\pi}{\hbar} |M_\pm|^2 g_{\rm 2D}^{2} \!\int_{0}^{\infty}\! f(E, E_{\rm F, \pm}) \left[1 - f(E, E_{\rm F, \mp}) \right] \, {\rm d}E, 
\end{align}
where $g_{\rm 2D} = m_{e} S / \pi \hbar^2$ is the 2D density of states,  $E_{\rm F,{\pm}} = E_{\rm F} \pm \Delta/2$ is the quasi-Fermi level of the $S_{z} = \pm 1/2$ electrons, $\Delta$ is the splitting of the quasi-Fermi levels, and $E = 0$ corresponds to the bottom of the conduction band.

Neglecting nuclear spin diffusion, we write the rate of change of the nuclear spin as
\begin{align} \label{a.eq:Iz_eq_korringa}
    \frac{{\rm d} \langle I_z \rangle}{{\rm d} t} = W_{+} - W_{-}.
\end{align}
Performing the integration in Eq.~(\ref{a.eq:flipRate}) we find
\begin{align} \label{a.eq:right_hand_side_pump}
    W_{+} - W_{-} =& \,\frac{\pi}{\hbar} k_{B} T \langle A^{2} \rangle \, \nu_{0}^{2} g_{\rm 2D}^{2} |\psi_{e, \rm free} (z)|^{4} \nonumber\\
    &\times \left[ \tanh \tfrac{\Delta}{2 k_{B} T} \left[I(I+1) - \langle I_z^2 \rangle \right] -  \langle I_z \rangle \right] \nonumber\\
    &\times f(\Delta, E_{\rm F}, T),
\end{align}
where 
\begin{align}
    f(\Delta, E_{\rm F}, T) =& \frac{1}{2} \ln{\left| \frac{e^{-E_{\rm F} / k_{B} T} + e^{\Delta / 2 k_{B} T}}{e^{-E_{\rm F} / k_{B} T} + e^{-\Delta / 2 k_{B} T}} \right|} \nonumber\\
    &\times \coth \tfrac{\Delta}{2 k_{B} T}
\end{align}
is a dimensionless factor that depends on the 2D electron gas density and temperature.
For a degenerate unpolarized gas $f(\Delta, E_{\rm F}, T) = 1$, and we recover the classic Korringa relaxation~\cite{korringa1950}.

In Eq.~(\ref{a.eq:right_hand_side_pump}), the term proportional to $\left[ I(I+1) - \langle I_{z}^{2} \rangle \right]$ can be identified as the pumping term, while the coefficient before $\langle I_{z} \rangle$ is the relaxation rate.
When the pumping beam is turned off after the optical cooling stage, the few resident electrons in Sample~B rapidly lose the light-induced polarization, and $P \to 0$ as $\Delta \to 0$.
We are then left with the Korringa-like nuclear spin relaxation rate
\begin{align} \label{a.eq:korringa_rel_rate}
    \Gamma_{\rm K}(z) =& \,\frac{\pi}{\hbar} \frac{k_{B} T}{1 + e^{-E_{\rm F} / k_{B} T}} \langle A^{2} \rangle \, \nu_{0}^{2} \, g_{\rm 2D}^{2} |\psi_{e, \rm free} (z)|^{4},
\end{align}
which recovers the Korringa behavior $\Gamma_{\rm K}(z) \propto k_{B} T$ at high electron densities, where $E_{\rm F} \gg k_{B} T$, but is spatially inhomogeneous due to the $|\psi_{e, \rm free} (z)|^{4}$ factor.

As explained in the main text, due to practical difficulties, we do not apply the full Eq.~(\ref{a.eq:diffusion_diffEq}) to describe the Sample~B data.
Instead, we use Eqs.~(\ref{eq:free_e_rate_eq}-\ref{eq:free_e_rate_eq_P_D_t}), which can be thought of as a spatial average of Eq.~(\ref{a.eq:diffusion_diffEq}).
There, we use the spatially averaged $\Gamma_{\rm K}^{\rm (avg)} = \frac{1}{L} \int_{\rm QW} \Gamma_{\rm K}(z) {\rm d} z$ as an estimate of the observed nuclear spin relaxation rate.
Considering that the Fermi energy and the electron density are linked by
\begin{align} \label{a.eq:density_formula}
    n_{\rm 2D} = \int_{0}^{\infty} g_{\rm 2D} f(E, E_{\rm F}) {\rm d}E,
\end{align}
we may directly plot $\Gamma_{\rm K}^{\rm (avg)}$ as a function of electron density using the parameters from Table~\ref{table:parameters}, see Fig.~\ref{fig:korringa_vs_density}.
The data presented in this Figure allow us to estimate the density of the electron gas puddles in Sample~B by varying the the diffusion contribution $\Gamma_{\rm D}$ to the total observed relaxation rate $\Gamma$, as described in the main text.
 
\bibliography{bibliography}

\end{document}